\documentclass{article}
\usepackage{ijcai-template/ijcai25}

\usepackage{times}
\usepackage{soul}
\usepackage{url}
\usepackage[hidelinks]{hyperref}
\usepackage[utf8]{inputenc}
\usepackage[small]{caption}
\usepackage{graphicx}
\usepackage{amsmath}
\usepackage{amsthm}
\usepackage{booktabs}
\usepackage{algorithm}
\usepackage{algorithmic}
\usepackage[switch]{lineno}

\usepackage{adjustbox}
\usepackage{booktabs}
\usepackage{comment}
\usepackage{xcolor}
\usepackage{amsmath}
\usepackage{amssymb}

\title{Decentralised multi-agent coordination for real-time railway traffic management}

\author{
Leo D'Amato$^{1,3}$
\and
Paola Pellegrini$^2$\and
Vito Trianni$^3$\\
\affiliations
$^1$Politecnico di Torino, DAUIN, Corso Castelfidardo, 34/d, Turin 10128, Italy.\\
$^2$Univ Gustave Eiffel, COSYS-ESTAS, F-59650 Villeneuve d’Ascq, France\\
$^3$Institute for Cognitive Sciences and Technologies, CNR, Via G. D. Romagnosi 18/A, 00196 Rome, Italy.\\
\emails
leo.damato@polito.it
}

\begin{document}

%%%%%%%%%%%%%%%%%%%%%%%%%%%%%%%%%%%%%%%%%%%%%%%%%%%%%%%%%%%%%%%%%%%%%%%%

\maketitle

%%%%%%%%%%%%%%%%%%%%%%%%%%%%%%%%%%%%%%%%%%%%%%%%%%%%%%%%%%%%%%%%%%%%%%%%

\begin{abstract}
    The real-time Railway Traffic Management Problem (rtRTMP) is a challenging optimisation problem in railway transportation. It involves the efficient management of train movements while minimising delay propagation caused by unforeseen perturbations due to, e.g,  temporary speed limitations or signal failures. This paper re-frames the rtRTMP as a multi-agent coordination problem and formalises it as a Distributed Constraint Optimisation Problem (DCOP) to explore its potential for decentralised solutions. We propose a novel coordination algorithm that extends the widely known Distributed Stochastic Algorithm (DSA), allowing trains to self-organise and resolve scheduling conflicts. The performance of our algorithm is compared to a classical DSA through extensive simulations on a synthetic dataset reproducing diverse problem configurations. Results show that our approach achieves significant improvements in solution quality and convergence speed, demonstrating its effectiveness and scalability in managing large-scale railway networks. Beyond the railway domain, this framework can have broader applicability in autonomous systems, such as self-driving vehicles or inter-satellite coordination.
\end{abstract}

%%%%%%%%%%%%%%%%%%%%%%%%%%%%%%%%%%%%%%%%%%%%%%%%%%%%%%%%%%%%%%%%%%%%%%%%

\section{Introduction}

In the realm of railway transportation, the \emph{real-time Railway Traffic Management Problem} (rtRTMP) \cite{10.1007/978-3-319-24264-4_45,PelMarPesRod15:ieeetits} is the problem of efficiently coordinating train movements across a railway network to counteract possible knock-on delays caused by traffic perturbations such as train malfunctions, signal failures, or temporary speed limitations. Knock-on delays are due to conflicts, that require external intervention on train paths through rerouting or rescheduling. Traditionally, such interventions have relied on centralised decision-making by human dispatchers, often with limited computational support, and primarily guided by personal experience.

In the academic literature, centralised approaches dominate, employing methods such as Integer Linear Programming (ILP) \cite{Caimi2012,Meng2014,TOLETTI2020100173}, Mixed Integer Linear Programming (MILP) \cite{Luan2020,Fischetti2017,PelMarPesRod15:ieeetits,Tornquist2007,LU2022102622,REYNOLDS2022105719,LeuBonCor:2023}, and graph-based formulations \cite{Corman2010,LamMan15,Mascis02,SamDarCorPar17,Bettinelli2017,Rod07}. However, centralised approaches struggle to scale with increasing network sizes due to computational constraints. These constraints may be overcome by developing decentralised approaches, where decision-making is distributed across individual trains acting as autonomous agents \cite{MarPel20:IEEE}. 

%The decentralised approach, termed the decentralised Railway Traffic Management Problem (dec-rtRTMP), eliminates the need for a central controller by allowing trains to adapt their schedules dynamically based on local interactions. This method scales well for large networks but sacrifices guaranteed optimality in favour of practical, effective solutions. Among existing approaches to the dec-rtRTMP, algorithms like Distributed Stochastic Algorithm (DSA) \cite{fitzpatrick2003distributed,Vanthielen2019}, multi-agent reinforcement learning \cite{Khadilkar,mohanty2020flatlandrl,flatland}, and swarm intelligence \cite{cui2017swarm} have shown promise but face challenges such as slow convergence or difficulty in handling complex scenarios.

Decentralisation potentially scales better to large-scale area networks, possibly at the cost of accepting non-optimal but still effective solutions.
One possible approach to the decentralised rtRTMP (dec-rtRTMP), presented by \cite{Vanthielen2019}, focuses on resolving conflicts individually by adjusting train schedules or routes. In contrast, \cite{shang2018distributed} suggest empowering trains to make individual decisions, optimizing their movements based on observations of preceding trains. Another proposal by \cite{cui2017swarm} introduces swarm intelligence, organizing trains into groups to address common conflicts collectively.
In addition, several methods from Artificial Intelligence (AI) have gained attention in this domain, especially following the recent developments in deep neural architectures \cite{JusupTrivellaCorman}. For instance, \cite{Khadilkar} advocates for reinforcement learning (RL), where agents learn from past experiences to make decisions. The Flatland challenges, initiated by European railway managers, have spurred research in this direction \cite{mohanty2020flatlandrl}, offering a simplified railway simulator for testing different machine learning (ML) approaches. However, deploying learning algorithms in such complex environment can be challenging and the lack of  guarantees on the feasibility of the obtained solution together with the black-box nature of these approaches, makes them difficult to accept by stakeholders.

%However, while AI offers promise, challenges remain. ML models for individual decision-making may be difficult to deploy without sufficient data for training.  Multi-agent RL models lack guarantees of always finding feasible solutions and can struggle in complex situations. Moreover, the black-box nature of deep neural approaches makes them difficult to understand and accept by stakeholders. 

An alternative approach consists in merging optimisation-based planning with self-organisation \cite{DAMATO2024100427}. In this approach, individual trains need to agree on possible schedules resulting from local optimisation by interacting with neighbours. 
\cite{DAMATO2024100427} proved the viability of the deployment to real world scenarios of such an hybrid approach to the dec-rtRTMP. More specifically, they studied a small portion of the line connecting Paris and Le Havre, in France, showing that, in case of traffic perturbations, not only the proposed approach is better than following the original timetable, but also it achieves performance comparable to the centralised state of the art. One of the key components is the self-organised process that enables reaching coordination among the agents on a feasible solution. To this end, decentralised consensus protocols \cite{Paxos:Lamport,amirkhani2022consensus} offer a framework for achieving such an agreement among agents without central control. Simple stochastic models, like the Voter Model, are often sufficient for a population to converge on shared opinions \cite{Holley:1975we}. These have been adapted to the dec-rtRTMP as a proof of concept \cite{DAMATO2024100427}, but without providing a clear problem formulation or a characterisation of the expected performance. In this study, we move a crucial step in this direction. 

We reframe the dec-rtRTMP as a Distributed Constraint Optimisation Problem (DCOP) \cite{fioretto2018distributed}, a mathematical framework used to model problems where multiple agents coordinate to find an optimal solution while respecting constraints imposed by their interactions. This perspective provides a principled foundation for designing and analysing decentralised coordination algorithms. Inspired by the literature on DCOP, we propose a novel decentralised multi-agent coordination algorithm that extends the classical Decetralised Stochastic Algorithm (DSA) \cite{fitzpatrick2003distributed}, tailored for solving the dec-rtRTMP. Our method leverages asynchronous local interactions among agents and employs adaptive strategies to efficiently resolve conflicts and optimise train routes and schedules.

Our contributions are summarised as follows:
\begin{itemize}\parsep0pt
    \item We provide a formal DCOP formulation of the dec-rtRTMP.
    \item We build a dataset for benchmarking decentralised solvers of the dec-rtRTMP.
    \item We introduce a novel stochastic algorithm for the solution of the dec-rtRTMP inspired by the DCOP solvers existing in the literature.
\end{itemize}

Despite being framed in the railway traffic management domain, the proposed approach is versatile and can be extended to other multi-agent systems requiring decentralised coordination, such as autonomous vehicle routing \cite{9078053} or inter-satellite coordination and scheduling \cite{picard:hal-03181968,YANG20214505}.

The rest of the paper is organised as follows: in Section \ref{sec:pf}, we describe our DCOP formulation of the dec-rtRTMP. In Section \ref{sec:consensus}, we present our decentralised multi-agent coordination algorithm adopted to solve a given instance of the dec-rtRTMP. In Section \ref{sec:exps}, we present the experimental settings we designed to study the main properties and the results obtained with our approach and compare them to those of a classical DSA algorithm. Finally, Section~\ref{sec:conclusions} concludes the paper with discussions about future research directions.

%%%%%%%%%%%%%%%%%%%%%%%%%%%%%%%%%%%%%%%%%%%%%%%%%%%%%%%%%%%%%%%%%%%%%%%%

\section{Problem Formulation}\label{sec:pf}

In this section, we first present the structure of a general DCOP and then we cast our dec-rtRTMP in terms of such framework.

\paragraph{General DCOP.} The main ingredients of a DCOP problem are agents and variables. Importantly, each variable is owned by an
agent; this is what makes the problem distributed. Formally, a DCOP is a tuple $\langle A, V, \mathfrak{D}, \mathcal{U}, \eta\rangle$, where:
\begin{itemize}

\item $A$ is the set of agents, $\left\{a_{1}, \ldots, a_{n}\right\}$.

\item $V$ is the set of variables, $\left\{v_{1}, v_{2}, \ldots, v_{m}\right\}$. In the most general formulation, one agent may control more than one variable, while some agents may control no variables at all (i.e. $n \neq m$). Here, we assume that each variable $v_i$ is controlled by exactly one agent $a_i$ ($n=m$).

\item $\mathfrak{D}$ is the set of variable-domains, $\left\{D_{1}, D_{2}, \ldots, D_{m}\right\}$, where each $D_{i} \in \mathfrak{D}$ is a finite set containing the possible values of variable $v_{i}$. 
%If $D_{i} \in \mathfrak{D}$ contains only two values (e.g. 0 or 1 ), then $v_{i}$ is called a binary variable.

\item A \emph{value assignment} is a pair $(v_i, d_i)$ where $d_{i} \in D_i$ denotes the value currently assigned to variable $v_i$ by the agent $a_i$. A \emph{partial assignment} $S$ is a set of value assignments $\{ (v, d): v \in W \subset V \}$ involving only a proper subset $W$ of the variable set $V$. A \emph{complete assignment} is a set of value assignments involving all the variables in $V$. We denote by $\mathfrak{S} = \mathfrak{S}_p \cup \mathfrak{S}_c$ the set of all possible assignments (both partial $\mathfrak{S}_p$ and complete $\mathfrak{S}_c$).

\item A set of constraints on partial assignments determine properties of value assignments and relations among them. Constraints can be \emph{unary} when involving only one variable, \emph{binary} when involving two variables or \emph{$k$-ary} when involving $k$ variables. In a DCOP formulation, constraints can be translated into cost functions (hence, the DCOP is a minimization problem) or into utility functions (hence, a maximisation problem). Here, we consider the latter option and introduce a set of functions $\mathcal{U} = \{u:\mathfrak{S} \rightarrow \mathbb{R}\}$, each providing the utility of satisfying a constraint on the assignment $S\in\mathfrak{S}$.

% \item $u: \mathfrak{S} \rightarrow \mathbb{R}$ is the \emph{utility function}. It maps every possible assignment to a utility score (the higher the better). Usually, only few assignments in $\mathfrak{S}$, referred to as \emph{constraints} of the problem, have a non-zero score. A constraint involving only one variable is called unary constraint, while a constraint involving two variables is called binary constraint.

\item The objective function $\eta: \mathfrak{S}_c \rightarrow \mathbb{R}$ maps each possible complete assignment to a score. For a complete assignment $S \in \mathfrak{S}_c$, we denote by $\mathfrak{S}(S,u)$ the set of all (partial) assignments that are subsets of $S$ and that have the cardinality required by the utility function $u$. Then the objective function is defined as:
$$
\eta(S) = \sum_{u\in\mathcal{U}}\sum_{S^{\prime} \in \mathfrak{S}(S,u)} u(S^{\prime}).%, \quad   \quad \forall S \in \mathfrak{S}_c
%\eta(S) = \sum_{S^{\prime} \in \mathfrak{S}(S)} u(S^{\prime}), \quad   \quad \forall S \in \mathfrak{S}_c
$$
\end{itemize}

The goal of the agents in a DCOP is to achieve an optimal solution, namely a complete assignment $S^*$ that maximises the objective function $\eta$. The search of an optimal solution is distributed because each agent can only access local information when assigning a value to the variable it controls. Indeed, each agent $a_i$ can interact with just a subset $N_i$ of the agents $A$. We refer to $N_i$ as the \emph{neighbourhood} of the agent $a_i$. This means that each agent $a_i$ can only observe the assignments of the variables controlled by its neighbours, i.e. the set $S(N_i) = \left\{ (v_j, d_j): a_j \in N_i \text{ and } v_j=d_j \right\}$, and thus it cannot fully evaluate the objective function $\eta$.

%This means that each agent $a_i$ can only communicate with a subset $N_i$ of the agents in the system. We refer to $N_i$ as the \emph{neighbourhood} of the agent $a_i$.

\paragraph{dec-rtRTMP as DCOP.} In a dec-rtRTMP, we can think of trains as agents. Each train $a_i$ can control its own path $v_i$, i.e. the sequence of track sections it is going to pass through in the next future with the time at which it will do so. At any time, each train can choose among a finite number of different paths between origin and destination, constituting the domain $D_i$ for the variable $v_i$. Thus, a value assignment $(v_i, d_i)$ corresponds to the decision of the agent $a_i$ to follow the path $d_i$.
%agent $a_i$ selecting the path $d_i$ as value for $v_i$. 
Each path $d_i$ is assigned a \emph{path utility} $u_r(v_i, d_i)$ by the agent $a_i$, i.e. a real number representing how convenient is for the train to follow such path. We assume this value to be normalised to be in the range $[0,1]$. The path utility represent a soft constraint on the validity and quality of a path, as its value can depend on the prediction of the delay accumulated at the end of the day by following the path $d_i$, the number of passengers carried and other factors. When selecting a path, trains should choose the one with the highest path utility but, at the same time, they must also account for the other trains in the system. More specifically, each agent $a_i$ has to select a path that not only possesses a high path utility, but that is also compatible with the paths of all its neighbours $N_i$. Two paths $d_i$ and $d_j$ belonging respectively to distinct agents $a_i$ and $a_j$ are said to be compatible if no track section is concurrently used by the trains when the two paths are simultaneously implemented. Compatibilities between two paths represent the binary constraints of the problem, and are associated to the compatibility utility $u_c$:
$$
u_c( (v_i, d_i), (v_j, d_j) ) = 
\begin{cases}
    1     & \text{if $d_i$ and $d_j$ are compatible}, \\
    0     & \text{otherwise}.
\end{cases}
$$
An optimal solution to the dec-rtRTMP is thus a complete assignment that maximises the sum of both utilities defined above, namely the objective function $\eta$.

Furthermore, since we only have unary and binary constraints, we can also provide a compact visual representation of a given instance of the dec-rtRTMP in the form of a pair of graphs $(\mathcal{G}_I, \mathcal{G}_C)$. The graph $\mathcal{G}_I$ is the \emph{interaction graph}, i.e., a graph whose nodes are agents and links represent neighbouring agents, while $\mathcal{G}_C$ is the constraint graph, i.e. a $n$-partite graph whose nodes are all the possible paths $d \in \bigcup_{D \in \mathfrak{D}} D$ generated by the agents and the links indicate the compatibility between paths. 

At this abstraction level, we do not discuss specific implementations of the path generation process, compatibility evaluation or neighbourhood identification since our focus is primarily on presenting a novel algorithm inspired by the DCOP literature to solve an instance of dec-rtRTMP. The interested reader can refer to \cite{DAMATO2024100427} for an example of specific implementation of similar components, presented in full details and in a real setting.

%%%%%%%%%%%%%%%%%%%%%%%%%%%%%%%%%%%%%%%%%%%%%%%%%%%%%%%%%%%%%%%%%%%%%%%%

% insert this in supplementary materials
% \input{src/centralised_sol}

%%%%%%%%%%%%%%%%%%%%%%%%%%%%%%%%%%%%%%%%%%%%%%%%%%%%%%%%%%%%%%%%%%%%%%%%

\section{A decentralised multi-agent coordination algorithm}\label{sec:consensus}

This section describes the self-organization process at the heart of our decentralised coordination approach. 
Inspired by the classical Decentralised Stochastic Algorithm (DSA) \cite{fitzpatrick2003distributed} from the DCOP literature, coordination is achieved through an iterative procedure, through which agents try to reach a (possibly optimal) solution to a given problem instance $(\mathcal{G}_I, \mathcal{G}_C)$ by only exploiting local information about their respective neighbours. 

At each iteration, the agent $a_i$ decides which value $d \in D_i$ to assign to the variable $v_i$ it controls. The algorithm starts with the agents performing a greedy assignment, i.e. $v_i=d^*$, where $d^* = \operatorname{argmax}_{d \in D_i} u_r(v_i,d)$. At each subsequent iteration, $a_i$ can decide either to keep its current assignment for $v_i$ or switch to another value $d^{\prime} \in D_i$. This choice depends solely on the local information available to the agent, that is, the path utility of the paths in $D_i$ and the degree of compatibility of $(v_i, d)$ with the current assignments $S(N_i)$ of its neighbours, $\forall d \in D_i$. 
Note that the path utility of neighbours' paths is not known, as this information is private to each agent.

More specifically, at each iteration $t$, the agent $a_i$, whose current assignment is $(v_i, d)$, operates the following steps:
\begin{enumerate}
    \item It randomly selects at most $k\geq1$ neighbours from $N_i$. We denote by $K_i(t)$ the subset of neighbours selected by the agent $a_i$ at time $t$. If $k\geq|N_i|$, then $K_i(t)=N_i$.
    \item It observes the current assignments of the selected neighbours, i.e. the set $S(K_i(t))$.  
    \item It creates a ranking over the set $D_i$. For each $d \in D_i$, the agent computes its rank $r(d,t)$ as the number of binary constraints satisfied by a potential assignment of $d$ to $v_i$ with respect to the current assignments $S(K_i(t))$ of the selected neighbours, i.e.     
    $$
    r(d,t) = \sum_{(v_j, d_j) \in S(K_i(t))} u_c \left((v_i, d),  (v_j, d_j)\right)
    $$
    %
    % \begin{align*} &r(d) =  \\ & \left| \left\{ (v_j, d_j) :  a_j \in K_i, v_j=d_j, u \left( \{ (v_i, d) , (v_j, d_j) \} \right) = 1 \right\} \right|. \\ \end{align*}
    %
    Hence, $r(d,t)$ represents the degree of compatibility of the value $d$ with the current assignments of the selected neighbours.
    \item It decides to keep its current assignment $(v_i, d)$ or to switch to a more compatible value $d^{\prime} \in D_i$ according to the following policy: 
    \begin{enumerate}
        \item if $r(d,t) = k$  (i.e. $d$ is compatible with the assignments of all the selected neighbours in $K_i(t)$), then the agent keeps $v_i=d$ as its current assignment.
        \item if $r(d,t) < k$, then the agent selects a more compatible value by sampling a value $d^{\prime}$ from the set of values $D_i$ satisfying the property $d^{\prime} = \operatorname{argmax}_{d \in D_i} r(d,t)$. In case of multiple values $d^{\prime}$ satisfying this property, a probabilistic choice is made with probability proportional to their respective path utility $u_r(v_i, d^{\prime})$.
    \end{enumerate}
\end{enumerate} 
The policy described in step 4 allows the agents to gradually adjust their assignments towards a configuration (solution to the problem) in which all neighbouring agents hold compatible values, while prioritising values with the highest possible utility score.
The parameter $k$ acts as a sort of learning rate for the algorithm.
With high values of the parameter $k$, the agent considers multiple neighbours during the decision making. This can lead the agent to seek compatibility with more neighbours at the same time, hence possibly increasing the speed of convergence towards a shared solution. Conversely, when $k$ is small (possibly, $k=1$) the agent only considers a few neighbours or just one at the time, and therefore the speed of convergence may be slower.
%\textcolor{blue}{
%[Qui mi avevi chiesto di spiegare l'effetto di K ma mi sembra un po' uno spoiler di quello che viene dopo nei risultati.]
%The speed of convergence to a solution is influenced by the hyperparameter $K$. When $K>1$, the agent considers multiple neighbours during the decision making and it can move faster towards a hypothesis that is compatible with the one selected by all its neighbours. The drawback is that agents with large value of $K$ can be trapped into deadlocks. This happens when the decision maker has an hypothesis that is compatible with most its neighbours but not all of them (hence such hypothesis is not part of any solution to the problem). In this scenario, the agent will select this hypothesis, influencing the decisions of all its neighbours at subsequent iterations. As a result, the system starts oscillating between configurations that are not solutions to the problem and it is not able to converge. Deadlocks never happen when $K=1$. In this case, the agent coordinates with one neighbour at a time and one incompatibility is enough to escape eventual deadlocks configurations, provided a sufficient number of iterations.
%}

\paragraph{Convergence Criteria.} 
We implemented an asynchronous multi-agent simulation to emulate real-world operation. Specifically, at each iteration, one single agent is selected randomly and updates its assignment following the default policy discussed above. This ensures that agents take turn with an approximate period of $n$ iterations.
This iterative process continues until one of the following conditions is met: (i) \emph{Convergence}: all agents achieve a state where their hypotheses are compatible with all their neighbours; (ii) \emph{Maximum Iterations}: a predefined maximum number of iterations is reached.

\paragraph{DSA.} Our algorithm differs from the classical DSA on three aspects:
\begin{itemize}
    \item A DSA agent \emph{always} interacts with all its neighbours (it cannot select a subset of them according to $k$).
    \item A DSA agent, with probability $1-\alpha$, can decide to keep it assignment a priori, regardless of the assignments of the neighbours. The parameter $\alpha$ is known as \emph{activation probability}.
    \item A DSA agent implements a different policy. In our implementation of DSA, the agent $a_i$ assigns a score $r(d)$ to each value $d \in D_i$ as follows:
    $$
    r(d) = u_r(v_i, d) + \sum_{(v_j, d_j) \in S(N_i)} u_c \left((v_i, d),  (v_j, d_j)\right)
    $$
    %given by the sum of the utilities of all the constraints (unary and binary) involving the assignment $(v_i, d)$. 
    Then, the agent greedily assigns to $v_i$ the value $d^*$ with the highest score $r$. In some variants of DSA \cite{Zhang_Wang_Xing_Wittenburg_2005}, the assignment can be $\epsilon$-greedy, i.e. with a (usually small) probability $\epsilon$, the agent assigns a random value $d$ to the variable $v_i$ instead of being always greedy. This helps escaping local maxima.
\end{itemize}
Hence, our algorithm improves over DSA by adding a more flexible interaction scheme and proposing a slightly different policy for the decision-making phase.

%%%%%%%%%%%%%%%%%%%%%%%%%%%%%%%%%%%%%%%%%%%%%%%%%%%%%%%%%%%%%%%%%%%%%%%%

\begin{figure*}[!t]
    \centering
    \includegraphics[width=\textwidth]{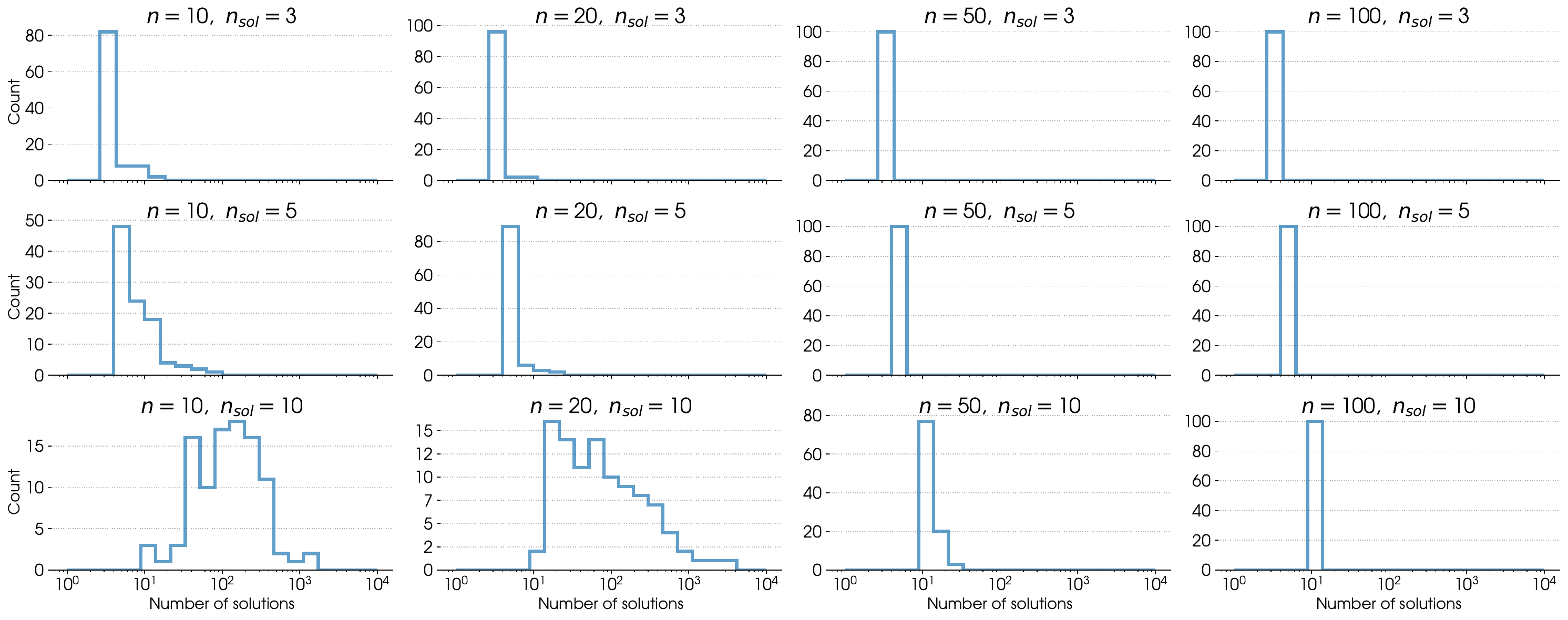}
    \caption{Distribution of the number of solutions per problem instance. Problem instances are grouped by the number of agents $n$ and by minimum number of solutions $n_{sol}$ we require the problem instance to have. For each combination of $n$ and $n_{sol}$, there are $100$ problem instances in our dataset. As evident from panels in the bottom left of the figure, when we add solutions to a graph $\mathcal{G}_C$ with few nodes, there is a high probability that the links of two solutions can be combined to form new solutions that we have not been explicitly inserted in the graph. This happens less frequently as the number of nodes in $\mathcal{G}_C$ increases.}
    \label{fig:sol_distr}
\end{figure*}

\section{Experiments}\label{sec:exps}

We conducted the evaluation of our algorithm on a dataset comprising synthetically generated problem instances defined by the following parameters: number of agents ($n$), interaction rate ($p_{\operatorname{int}}$), maximum number of paths per agent ($n_d$), minimum number of solutions ($n_{sol}$), and a random seed ($\sigma$). The interaction graph $\mathcal{G}_I$ is created as a connected graph with $n$ nodes and edges added with probability $p_{\operatorname{int}}$. Each agent in $\mathcal{G}_I$ generates between 1 and $n_d$ paths, forming the nodes of the constraint graph $\mathcal{G}_C$. Paths have path utilities of 1 (preferred) or 0.1 (less preferred), following a standard practice for experimentation in multi-alternative decisions in which the less-preferred options are considered distractors with an equally low utility \cite{Reina:2017jl}. Edges in $\mathcal{G}_C$ are initialized to construct $n_{sol}$ solutions and refined to ensure that each node has at least degree $1$. While $n_{sol}$ sets a minimum, the total number of solutions is computed post-generation, as described in  Section S3 of the Supplementary Material. The distribution of the total number of solutions of the problem instances in our dataset is reported in Figure \ref{fig:sol_distr}. Problem instances were generated with fixed $p_{\operatorname{int}}=0.3$ and $n_d=8$, varying $n \in \{10, 20, 50, 100\}$ and $n_{sol} \in \{3, 5, 10\}$, yielding 1200 instances. Further details about the data generation process can be found in section S2 of the Supplementary Material. The dataset will be publicly available with the aim of stimulating research on dec-rtRTMP.

%We conducted the evaluation of our algorithm on a synthetic dataset consisting of 1200 distinct problem instances. 
%In Supplementary Material S2 we detail how we generated the dataset and in Supplementary Material \ref{sec:data_gen} how we computed the set of all solutions in a separate step after the generation.
%Note that each problem instance is required to have at least $n_{sol}$ solutions by design. However, by design, we do not control the total number of solutions in a problem instance. 

We benchmark our algorithm against a classical DSA (with $\alpha = 0.9$ and $\epsilon = 0$, but see the Supplementary Matrial for additional values of $\alpha$) in terms of quality of the solution and speed of convergence. Below, we present the results of the execution of our algorithm. 

%%%%%%%%%%%%%%%%%%%%%%%%%%%%%%%%%%%%%

\subsection{Results}\label{subsec:res}

%On each problem instance we tested three types of agents:
%%Each problem instance was subjected to 300 executions of our algorithm, encompassing three different types of agents. For each combination of agent type and problem instance, our algorithm was executed 100 times with unique seeds to ensure robustness. The tested agent types include:
%
%\begin{enumerate}
%\item $\texttt{k\_1}$: Each agent in the system considers only one neighbor during the decision-making process.
%\item $\texttt{k\_all}$: Each agent in the system considers all its neighbors during the decision-making process.
%\item $\texttt{k\_ada}$: Each agent in the system initially considers all its neighbors during the decision-making %process and, after $1000$ iterations, it starts reducing the number of neighbors considered over iterations, until it considers only a single neighbor ($K=1$).
%\end{enumerate}
We exploited the synthetic dataset to test the quality of the proposed decentralised multi-agent coordination algorithm varying the value of the parameter $k$. In more detail, we deployed agents that consider only one neighbour during the decision-making process (hence, $k=1$, hereafter referred to as $\texttt{k\_1}$), as well as agents that consider all their neighbours during the decision-making process (hence, $k=\infty$, hereafter referred to as $\texttt{k\_all}$). 
On each problem instance, we executed our decentralised algorithm $100$ times per type of agent, every time using a different random seed to ensure robustness and reproducibility. On each execution, we set an upper bound on the number of iterations. The upper bound has been fixed to $10^5$ and every time our algorithm exceeds this bound the execution is interrupted resulting in a failure.
The same experimental setting holds for the classical DSA (hereafter referred to as $\texttt{dsa}$).

%We evaluated our approach in terms of quality of the solution found at the end of the self-organisation process and in terms of speed of convergence to such solution. For what concerns the solution quality, for each problem instance we created a ranking of all its solutions, from the optimal ones (the complete assignments maximising the objective $\eta$) to the ones with smallest value of $\eta$, and we measured at which level of such ranking the solution found by our algorithm is located (on average, over 100 runs). For the speed of convergence, we measure the time taken by the $n$ agents to converge to a solution. Note that the default policy implies that any solution represents an absorbing state for the system, that is, no agent changes state any more once a solution is found.

We evaluated our approach based on the quality of the solution obtained at the end of the self-organization process and the speed of convergence to that solution. Regarding solution quality, we ranked all solutions for each problem instance in decreasing order of their objective value $\eta$, with the optimal solutions (those maximizing $\eta$) at the top. We then assessed the position of the solution found by our algorithm within this ranking, averaging the results over 100 runs. For speed of convergence, we measured the number of steps required for the $n$ agents to converge to a solution. It is important to note that under the policy described in Section \ref{sec:consensus}, any solution acts as an absorbing state for the system, meaning no agent changes state once a solution is found.

\begin{figure*}[!t]
    \centering
    \includegraphics[width=\textwidth]{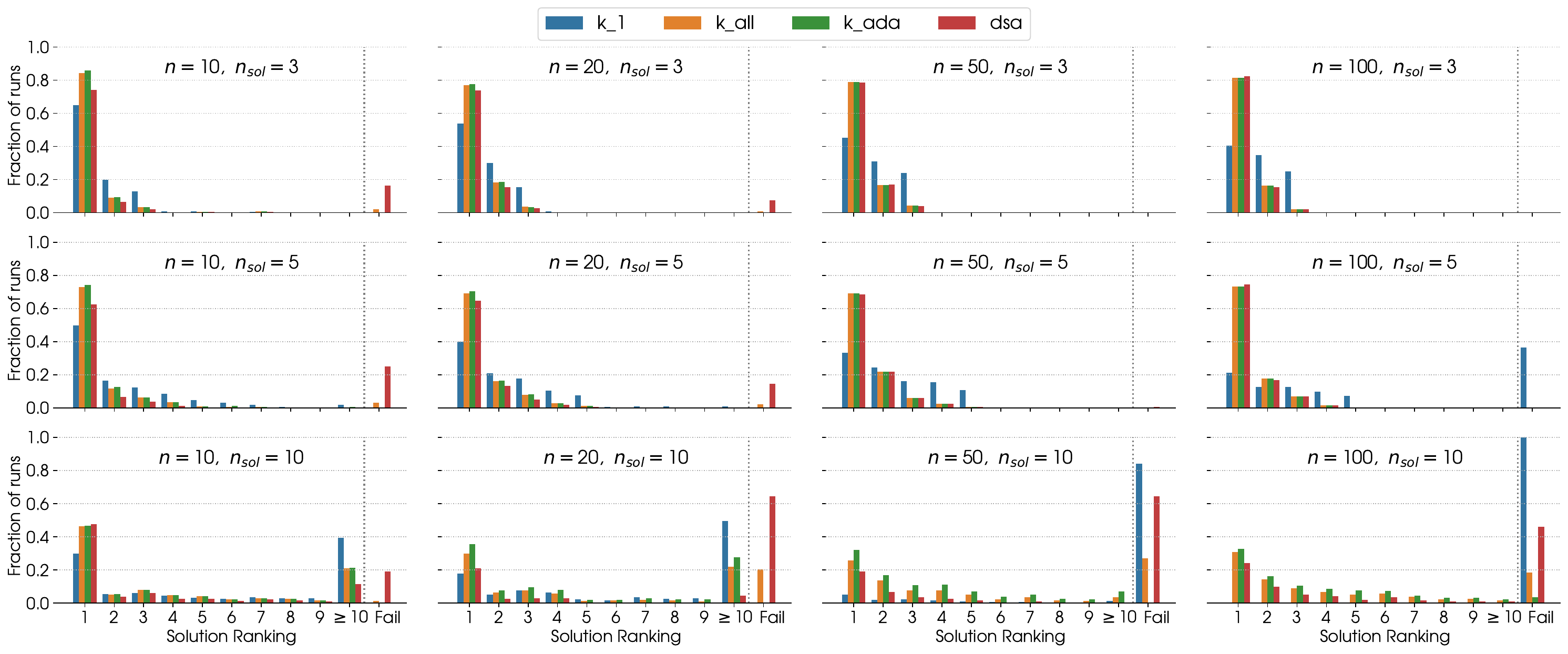}
    \caption{Ranking of the solutions given by our algorithm. Data are grouped by the type of agent, by number of agents $n$ and by minimum number of solutions $n_{sol}$ we require the problem instance to have. Bars with label ``$1$'' represent the fraction of executions that converged to an optimal solution. Bars with label ``$\geq 10$'' represent the fraction of executions that converged to a solution in position greater than 10 in the ranking. Bars with label ``Fail'' represent the fraction of executions that exceeded the upper bound of $10^5$ iterations.}
    \label{fig:ranking}
\end{figure*}

Figure~\ref{fig:ranking} shows the average position in the ranking of the solution given by the proposed approach compared to DSA. Additionally, it shows the fraction of runs that end up in a failure (see the last bar in each plot labelled `Fail'). We first notice that $\texttt{k\_1}$ only fails on large problem instances (large $n$ and $n_{sol}$), while $\texttt{k\_all}$ presents a small number of failures in many conditions, even for small problem instances ($n=10$, $n_{sol}=3$).
%We first notice that, in all conditions, $\texttt{k\_1}$ never fails, while $\texttt{k\_all}$ presents in nearly any condition a small number of failures, even for small problem instances ($N=10$, $\bar{S}=3$).
By analysing the behaviour of the system in such instances, we discovered that the large value of $k$ leads to deadlocks, whereby the system oscillates between configurations that are not solutions to the problem.
%This happens when the decision maker has an hypothesis that is compatible with most of its neighbours, but not all of them (hence, such hypothesis is not part of any solution to the problem). In this scenario, the agent will select this hypothesis, influencing the decisions of all its neighbours at subsequent iterations. As a result, the system starts oscillating between configurations that are not solutions to the problem and it is not able to converge. 
We observed deadlocks with any value of $k>1$. Instead, with $k=1$, deadlocks are not possible, because any agent interacts with only a single neighbour and finding one incompatibility is enough to escape eventual deadlock configurations, provided that a sufficient number of iterations is performed.

The discovery of deadlocks lead us to introduce an adaptive strategy (referred to as $\texttt{k\_ada}$), in which each agent in the system initially considers all its neighbours during the decision-making process and, after $1000$ iterations, it starts reducing the number of neighbours considered, linearly over a window of $10^4$ iterations, until it considers only a single neighbour ($k=1$).
%it starts reducing the number of neighbors considered over iterations, until it considers only a single neighbor ($K=1$). The hyperparameter $K$ decreases linearly over a window of $10^4$ iterations. 
This will remove deadlocks as soon as agents reach the value $k=1$. At the same time, it can preserve the features of high values of $k$ such as an expected higher convergence rate, as conjectured in Section~\ref{sec:consensus}. In Figure~\ref{fig:ranking}, we show that $\texttt{k\_ada}$ always reaches a feasible solution but for very large problem instances ($n=100$, $n_{sol}=10$), where failures are just a few and they are not due to deadlocks. Hence, the adaptive approach successfully removes the deadlock occurrence issue.
% As already specified in Section~\ref{sec:consensus}, having $K>1$ improves the speed of convergence our algorithm at the price of possibly getting trapped into deadlocks while having $K=1$ ensure convergence but it takes longer to achieve it. We introduce the adaptive agent to have the best of both worlds. As confirmed by results in Section~\ref{subsec:res}, a system made of adaptive agent rapidly converges to a solution and in case of deadlocks it is able to escape them by falling back to a system made of $K=1$ agents.

\begin{figure*}[!t]
    \centering
    \includegraphics[width=\textwidth]{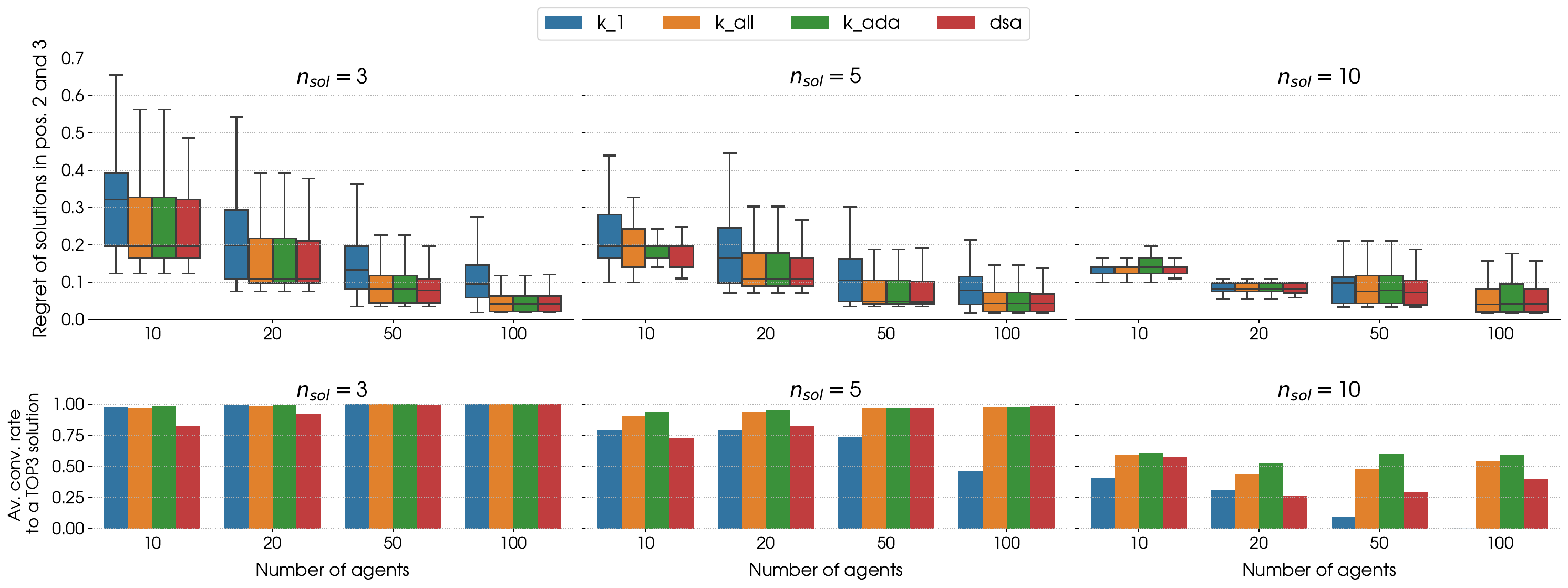}
    \caption{Top: Regret (percentage loss from the optimal solution value) of the solutions given by our algorithm in position 2 or 3 of the ranking. Data are grouped by type of agent, by number of agents $n$ and by minimum number of solutions $n_{sol}$ we require the problem instance to have.
    Bottom: Average convergence rate to a top-3 solution.}
    \label{fig:regret}
\end{figure*}

Looking at the ranking of the produced solutions in Figure~\ref{fig:ranking}, we found that, when the agents consider all their neighbours during the decision making phase, the rate of convergence to an optimal solution is not significantly impacted by the number of agents in the system. As an example, consider the first row in Figure~\ref{fig:ranking}: the $\texttt{k\_all}$ and  $\texttt{k\_ada}$ strategies have a convergence rate to an optimal solution around $80\%$, regardless of the number of agents in the system. The same does not hold true for $\texttt{k\_1}$, as the convergence rate to the optimal solution drops as soon as the number of agents increases. 

The second factor impacting on the rate of convergence to an optimal solution is the minimum number of solutions of the problem instance. As shown in Figure~\ref{fig:ranking}, the larger $n_{sol}$, the harder it is for our algorithm to converge to an optimal solution, regardless of the type of strategy adopted and of the number of agents in the system. Additionally, we note that instances that have many possible solutions are more difficult to solve (e.g., the rank of the solutions found is generally lower when $n=20$ and $n_{sol}=10$, where there are instances with a large amount of solutions, see Figure~\ref{fig:sol_distr}). In such conditions, most solutions have the same score $\eta$ due to the way in which solutions are constructed. Hence, agents do not particularly favour one solution over the other, often ending up in lower-ranked solutions. 

It is important to note that, when our algorithm does not converge to an optimal solution, it usually converges to a solution in the top-3 of the ranking, as reported by the bottom panel of Figure~\ref{fig:regret}. For example, the adaptive strategy, even on the most difficult problems, converges to a top-3 solution at least $50\%$ of the times. However, this is not enough to properly measure the quality of our algorithm, since it is not obvious if the value of a solution in second or third position in the ranking is close to the optimal value. For this reason, for each solution obtained with our algorithm, we measure the percentage deviation of its score $\eta$ from the score of the optimal solutions (first position in ranking), namely the regret. The upper panel of Figure~\ref{fig:regret} shows the regret of solutions in second and third positions of the ranking. We can observe that $\texttt{k\_all}$ and $\texttt{k\_ada}$ have similar performance, with a median regret up to $10\%$ on most executions and around $20\%$ in the worst case, and they clearly outperform the  $\texttt{k\_1}$ strategy. This means that, even when it is not optimal, the quality of the solution of our algorithm is largely acceptable.

\begin{figure*}[!t]
    \centering
    \includegraphics[width=\textwidth]{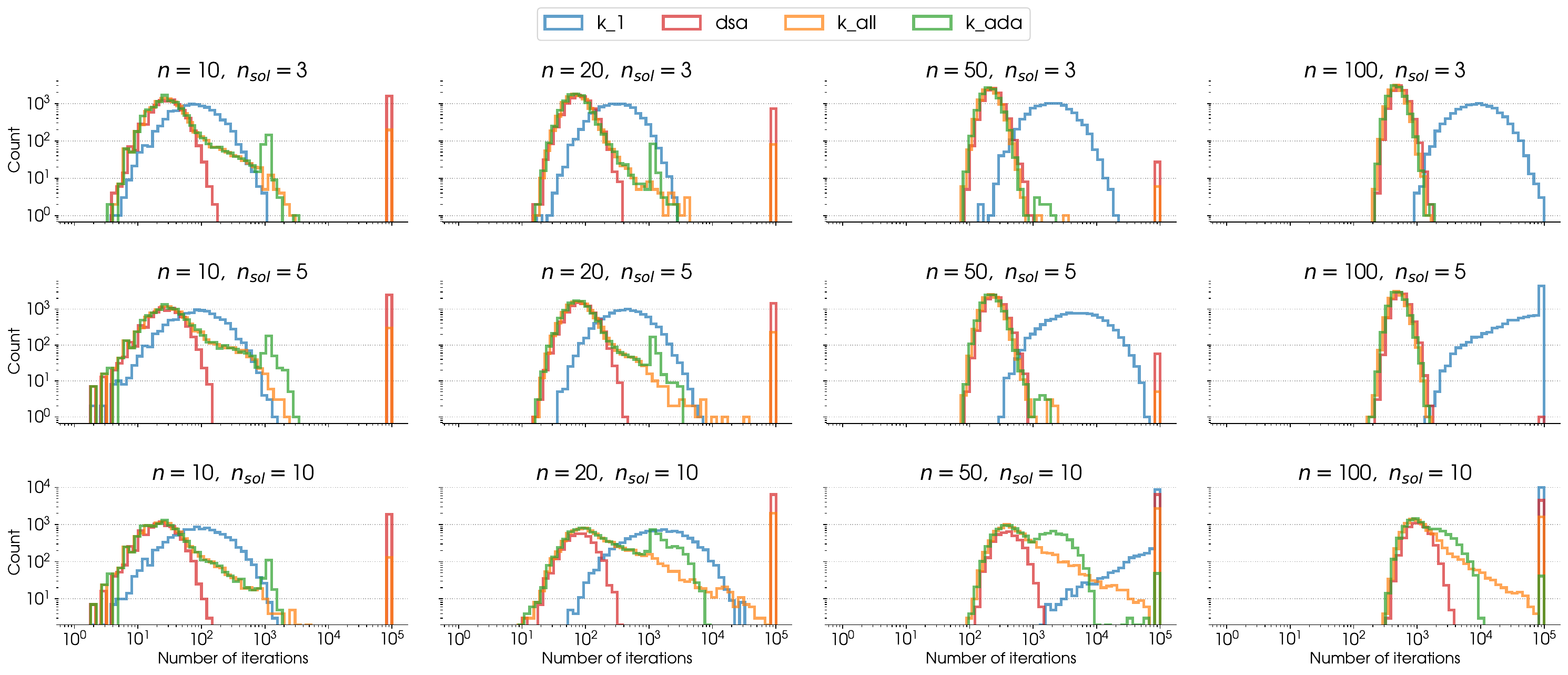}
    \caption{Distribution of convergence times (in terms of number of iterations) per type of agent on problem instances grouped by number of agents $n$ and by minimum number of solutions $n_{sol}$ we require the problem instance to have. For each of the 100 problem instances characterised by $n$ and $n_{sol}$, we performed $100$ executions of our algorithm for each agent type. Each execution has an upper bound of $10^5$ iterations, beyond which it fails.}
    \label{fig:convTime}
\end{figure*}

With respect to the speed of convergence, Figure~\ref{fig:convTime} demonstrates that the $\texttt{k\_ada}$ strategy is the best performing one. The $\texttt{k\_all}$ strategy shows similar performance but it fails on many problem instances, even those with a small number of agents and few solutions due to deadlocks. 
% A closer examination of this cases revealed that agent with $K>1$ can be trapped into deadlocks. More specifically, we observed that in this cases the system starts oscillating between two configurations, none of them being a solution, and it is not able to escape. 
The $\texttt{k\_ada}$ strategy does not suffer from deadlocks, since, after a certain number of iterations, $k$ falls back to 1 and the agents are able to escape the deadlock. This is evident from the bumps in the tails in the histograms of $\texttt{k\_ada}$ in Figure~\ref{fig:convTime}, which correspond to those problem instances on which $\texttt{k\_all}$ failed to converge and that are instead solved by $\texttt{k\_ada}$ by lowering the value of $k$. On the other hand, when $k=1$, our algorithm never experiences deadlocks but is much slower to converge to a solution, to the point that in problem instances with a lot of agents and a lot of solutions, it never converges before the threshold of $10^5$ iterations.
%%%%%%%%%%%%%%%%%%%%%%%%%%%%%%%%%%%%%

For what concerns $\texttt{dsa}$, it shows slightly worse performance with respect to $\texttt{k\_all}$ in terms of solution quality (see Figure \ref{fig:ranking}). In fact the two strategies are similar since each agent considers all its neighbours during the decision making phase. However, the policy implemented by the $\texttt{dsa}$ agents results in larger convergence times and also in several convergence failures (see Figure \ref{fig:convTime} and the Supplementary Material for other values of $\alpha$).
%However, since the $\texttt{dsa}$ agents implement a dummy policy with respect to the policy described in \ref{sec:consensus}, the converge time is much longer resulting in a lot of failures (see Figure \ref{fig:convTime}).

In conclusion, $\texttt{k\_ada}$ seems to be the best algorithm since it features a speed of convergence similar to $\texttt{k\_all}$ but, as $\texttt{k\_1}$, it is not subject to deadlocks. Also, it achieves a good rate of convergence to a top-3 solution of any given problem instance.

\begin{comment}
main findings:
\begin{itemize}
    \item the rate of convergence to the optimal solution if not significantly impacted by the size of the problem instance (number of agents) when k is close to the number of neighbors of each agent. When k=1 we can observe a significant decrease in the performance.
    \item The number of solutions in a problem instance significantly impacts the rate of convergence to the optimal solution of all the agent types. The more the solutions, the harder is to converge to the optimal one
    \item Even though the algorithm do not converge to an optimal solution, it converges to something that is close to it (see the regret). This holds for all the agents
    \item larger value of k significantly speed up the convergence of the algorithm (less iteration are needed to find a solution). Drawback: if k is large and the problem instance is small we can have deadlocks. The adaptive agent solves this problem.
\end{itemize}
\end{comment}

%%%%%%%%%%%%%%%%%%%%%%%%%%%%%%%%%%%%%%%%%%%%%%%%%%%%%%%%%%%%%%%%%%%%%%%%

\section{Conclusions}
\label{sec:conclusions}

In this paper, we presented a decentralised multi-agent coordination algorithm to solve the dec-rtRTMP, a challenging optimisation problem in railway transportation. It involves the efficient management of train movements on a railway network while minimising delay propagation caused by unexpected events such as temporary speed limitations or signal failures. 
Our contributions include a formal DCOP formulation of the dec-rtRTMP, a benchmark dataset for evaluating decentralised solvers, and a novel stochastic algorithm for decentralised multi-agent coordination. Through extensive experimentation, we tested three variants of our approach against a classical DSA from DCOP literature and we evaluated the results in terms of solution quality and convergence speed. We found that our algorithm, when using adaptive agents, achieves high-quality solutions, typically ranking within the top-3 solutions, significantly outperforming the classical DSA. Additionally, the algorithm demonstrates robustness in convergence, particularly in scenarios with varying numbers of agents and solution complexity.

Our work builds upon recent research trends that advocate for decentralisation in railway traffic management. Our DCOP-based reformulation of the dec-rtRTMP represents a proposal for an abstract mathematical framework to deal with such problem. However, translating this abstract framework into concrete case studies can be challenging. Depending on the application, one should carefully define some key aspects like the concept of interaction between trains, the concept of compatibility between paths of distinct trains and other operational aspects. Nonetheless, recent work on railway traffic management has demonstrated that similar approaches to decentralised railway traffic management are feasible \cite{mohanty2020flatlandrl,DAMATO2024100427}, paving the way for new developments in this field.

Besides addressing the dec-rtRTMP, the proposed algorithm for DCOP could be applied to other application domains, related to traffic management or to other decentralised coordination problems \cite{9078053,10.5555/3535850.3535969,10.1613/jair.1.16997,picard:hal-03181968,YANG20214505}. The key feature of target problems are the decentralised choice among a set of alternatives, respecting the compatibility of choices among neighbouring agents. A key aspect of the proposed approach is a policy that prioritises binary constraints (e.g., compatibility of value assignments) over unary constraints (e.g., quality of value assignment), as the former determines the ranking of value assignments while the latter is exploited only for choosing among equally-ranked assignments. On the contrary, DSA merges all constraints in a single utility.
%, and we conjecture that this could be a penalising factor resulting in slower convergence, because high utility of respecting the unary constraints may mask the low utility of the binary ones. 
In future work, we will deepen our analyses to understand to what extent the prioritisation of binary constraints over unary ones is beneficial to convergence or detrimental to quality. Additionally, we will formally address the existence of deadlock conditions, to find ways of avoiding them while maximising the coordination ability within a neighbourhood of agents. Finally, we aim at deploying adaptive algorithms that learn the parameters from the outcome of previous coordination rounds.    

%%%%%%%%%%%%%%%%%%%%%%%%%%%%%%%%%%%%%%%%%%%%%%%%%%%%%%%%%%%%%%%%%%%%%%%%

%\section*{Acknowledgments}
%The work presented in this paper has been carried out in the context of the SORTEDMOBILITY project. This project is supported by the European Commission and funded under the Horizon 2020 ERA-NET Cofund scheme under grant agreement N 875022. Vito Trianni and Leo D'Amato acknowledge partial support by TAILOR, a project funded by EU Horizon 2020 research and innovation program under GA No 952215.

%%%%%%%%%%%%%%%%%%%%%%%%%%%%%%%%%%%%%%%%%%%%%%%%%%%%%%%%%%%%%%%%%%%%%%%%

\bibliographystyle{ijcai-template/named}
\bibliography{src/references}

%%%%%%%%%%%%%%%%%%%%%%%%%%%%%%%%%%%%%%%%%%%%%%%%%%%%%%%%%%%%%%%%%%%%%%%%

\newpage
\appendix
%reset the counter
\setcounter{figure}{0} 
\renewcommand{\thefigure}{S\arabic{figure}}
\renewcommand{\theHfigure}{S\arabic{figure}}

\setcounter{section}{0} 
\renewcommand{\thesection}{S\arabic{section}}
% Custom appendix autoref labels
\preto\appendix{%
   \renewcommand{\sectionautorefname}{Appendix}%
   \renewcommand{\subsectionautorefname}{Appendix Subsection}%
}

\begin{center}
    \makebox[1\width]{
    \begin{tabular}{c}
        \huge \textbf{Supplementary Materials}
    \end{tabular}
    }
\end{center}

\section{Introduction}

The structure of supplementary material is as follows:
Section \ref{sec:data_gen} illustrates the generation process of a problem instance.
Section~\ref{sec:centralised_appr} describes an integer linear-programming formulation of the problem, useful to compute optimal solutions in a centralised way, as a benchmark for decentralised approaches.
Section \ref{sec:suppl_dsa} shows the effect of the activation probability $\alpha$ by running experiments with $\alpha=1$, $0.9$ and $0.7$.

\begin{figure}[ht!]
    \centering
    \includegraphics[width=1\linewidth]{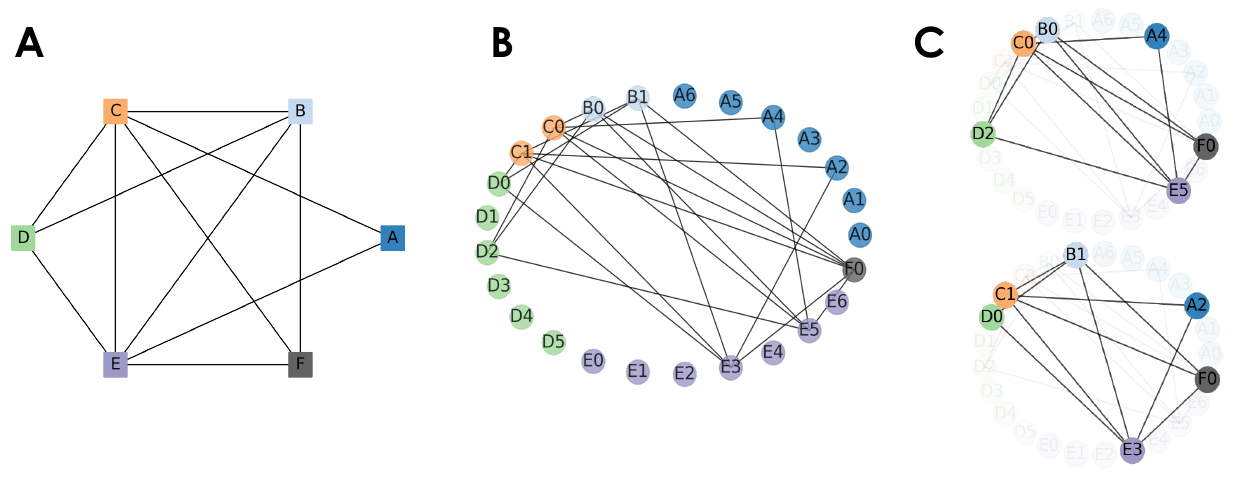}
    \caption{An example of problem instance in our DCOP-based formulation of the dec-rtRTMP. 
        \textbf{(A)} Interaction graph $\mathcal{G}_I$.
        \textbf{(B)} Constraint graph $\mathcal{G}_C$.
        \textbf{(C)} Two possible solutions of the problem instance.
        }
    \label{fig:pi_example}
\end{figure}

Recall from the main text that an instance of dec-rtRTMP can be compactly represented as a pair of graphs $(\mathcal{G}_I, \mathcal{G}_C)$ where $\mathcal{G}_I = (\mathcal{V}_I, \mathcal{E}_I)$ is the \emph{interaction graph}, i.e. a graph whose nodes $\mathcal{V}_I$ are agents and links $\mathcal{E}_I$ represents neighbouring agents, and $\mathcal{G}_C = (\mathcal{V}_C, \mathcal{E}_C)$ is the constraint graph, i.e. a $n$-partite graph whose nodes $\mathcal{V}_C$ are all the possible paths $d \in \bigcup_{D \in \mathfrak{D}} D$ generated by the agents and the links $\mathcal{E}_C$ denote compatible paths. An example of problem instance is reported in Figure \ref{fig:pi_example}.

%Since a path is an hypothesis formulated by the train about future evolution of railway traffic, from here on we refer to it simply as \emph{hypothesis}.

\section{Data Generation}\label{sec:data_gen}

% The dataset contains synthetically generated problem instances. 
To generate a problem instance, we specify (i) the number of agents $n$, (ii) the interaction rate between agents $p_{\operatorname{int}}$, (iii) the maximum number of paths each agent can generate $n_d$, (iv) the minimum number of solutions $n_{sol}$ we want the problem instance to have, and (v) a random seed $\sigma$ since the generation process is not deterministic.

The parameters $n$ and $p_{\operatorname{int}}$ determine the stochastic generation of the interaction graph $\mathcal{G}_I$: $n$ corresponds to the number of nodes of $\mathcal{V}_I$ while $p_{\operatorname{int}}$ corresponds to the probability of an edge between two nodes, and therefore it varies between 0 and 1. $\mathcal{G}_I$ is generated as a connected graph because, when it is not connected, our algorithm can be applied to each connected component independently. To generate $\mathcal{G}_I$ as random connected graph, we operate the following steps: we first create a tree with nodes $\mathcal{V}_I$ and then we add random links between nodes with probability $p_{\operatorname{int}}$.
%is the same as the one for generating a Erd\H{o}s-R\'enyi graph \cite{Erdos:1959:pmd}, except that we enforce the resulting graph to be connected by adding one random edge for each node before adding any other edge.

The parameters $n_d$ and $n_{sol}$ are instead used to generate the constraint graph $\mathcal{G}_C$. Each agent in $\mathcal{G}_I$ generates a random number of paths between $1$ and $n_d$. Each path $d$ is associated with a utility value $u(d)$. We assume that each agent has one preferred path and other less desirable ones. The former has a utility value equal to 1, while the latter have a small utility value equal to 0.1. This choice follows a standard practice for experimentation in multi-alternative decisions in which the less-preferred options are considered distractors with an equally low value~\cite{Reina:2017jl}.
%Possible values for $u_h$ are either $0.1$ or $1$, with the constraint that exactly one path among the generated ones has a value of $1$. The rationale behind this design choice...  } 
The set of paths of all the agents correspond to the nodes $\mathcal{V}_C$ of $\mathcal{G}_C$.
The links in $\mathcal{G}_C$ are inserted according to the following procedure: we first construct $n_{sol}$ different solutions (see below), and the links constituting such solutions are then added to $\mathcal{G}_C$. At this point not all the nodes in $\mathcal{G}_C$ have a link and thus we randomly add one link for each node with degree $0$ to avoid having paths that are not compatible with any other path. When adding random links, we must take into account the interactions in $\mathcal{G}_I$. Indeed, if two agents do not interact, there is no reason to check the compatibility of their paths and thus we only randomly add links between paths of neighbouring agents.

The construction of a solution to the problem is straightforward: it is enough to randomly select one path per agent and then add all the possible links between paths of neighbouring agents in $\mathcal{G}_C$. Note that even if we add $n_{sol}$ solutions to a problem instance, the total number of solutions can be greater than $n_{sol}$ because the union of the links constituting two solutions can generate several additional solutions, especially if the number of nodes in $\mathcal{V}_C$ is small. This means that the total number of solutions in a problem instance cannot be controlled at generation time but must be computed after the generation, in a centralised fashion, by means of the CPLEX solver, as described in Section~\ref{sec:centralised_appr}. Figure 1 in the main text shows the distribution of the number of solutions in the problem instances of our dataset.  Note that 
%$n_{sol}$ determines the minimum number of solutions, but not the total number, as shown in Figure~\ref{fig:sol_distr}. Indeed, 
the highest number of solutions is found in instances in which $n$ is not much larger than $n_{sol}$, with instances in the group determined by $n=20$ and $n_{sol}=10$ having up to $10^4$ possible solutions. Instead, when $n$ is much larger than $n_{sol}$ (e.g., $n=100$ and $n_{sol}=10$), only the minimum number of solutions $n_{sol}$ is present.

We generated problem instances by fixing the interaction rate $p_{\operatorname{int}}=0.3$ and the maximum number of paths per train $n_d=8$. These values have been selected empirically to obtain a sufficiently rich topology for $\mathcal{G}_I$ and $\mathcal{G}_C$. Then, we generate problem instances by varying $n \in \{10, 20, 50, 100\}$ and $n_{sol}\in\{3, 5, 10\}$. For each combination, we generate 100 instances by varying the random seed $\sigma\in\{0,\ldots,99\}$. The seed $\sigma$ is  used to make the generation process reproducible. Overall, the total number of problem instances in our dataset is $1200$.\footnote{The dataset will be publicly available with the aim of stimulating research on dec-rtRTMP.}

\section{Centralised solution to the coordination problem}\label{sec:centralised_appr}

In this section, we aim at determining if a given problem instance $(\mathcal{G}_I, \mathcal{G}_C)$ admits a solution and, in case it does, we want to compute all possible solutions while identifying the optimal ones.
%\textcolor{blue}{forse conviene specificare che non è possibile farlo in maniera decentralizzata ? }
To accomplish this, we can reformulate the problem as an integer linear programming (ILP) task and utilize widely available software such as CPLEX \footnote{https://www.ibm.com/it-it/products/ilog-cplex-optimization-studio} as solvers.
%
%Given a problem instance $(\mathcal{G}_I, \mathcal{G}_C)$, we would like to know whether it admits a solution or not and, in case it does, what are all the possible solutions and which ones are optimal. A method to achieve this goal is by reformulating it as an integer linear programming (ILP) problem and solve it by means of standard software like CPLEX. 
%
%The output of our decentralised algorithm at convergence is a solution to the problem instance being solved. A way to assess the performance of the our algorithm is to compare the value of the solution found with the value of the optimal solution to the problem being solved. A method to compute the optimal solution of a problem instance is by reformulating it as an integer linear programming (ILP) problem and solve it by means of tools like CPLEX. 
%
Such a centralised approach provides a reference for the evaluation of the decentralised solutions in our experiments, as discussed in the main text. 

%The output of our decentralised algorithm at convergence is a solution to the problem instance being solved. In order to assess the performance of the consensus procedure we must access all the solutions of the problem instance. One way to compute the entire set of solutions of a problem instance is by reformulating the rtRTMP as an integer linear programming (ILP) problem and solve it by means of a solver like CPLEX. Such an approach is no longer decentralised and serves as a baseline for our experiments in section \ref{sec:exps}. 
To define the ILP forumulation, we first introduce the mapping $\alpha : \mathcal{V}_C \mapsto \{1, \ldots, n\}$ associating to each path $d$ the index of the agent it belongs to. Then, we define the binary decision variables $y_d$ as follows:
\begin{equation}
\label{eq:binary}
y_d = \left\{
                \begin{array}{ll}
                  1 \quad \mbox{if $v_{\alpha(d)}=d$}     \\
                  0	\quad \mbox{otherwise}\\
                \end{array}
                 \right. \quad \forall d \in \mathcal{V}_C
\end{equation}
where the notation $v_{\alpha(d)}=d$ means that the path $d$ has been assigned to the variable $v_{\alpha(d)}$ controlled by the agent $a_{\alpha(d)}$. 
Recall also that each node $d \in \mathcal{V}_C$ has a value $u(d)$. 
Then, the centralized optimization problem consists in finding the appropriate complete assignment of paths to maximise the following objective function: 
\begin{equation}\label{eq:objective}
\max\sum_{d \in \mathcal{V}_C} u(d) \ y_d 
\end{equation}
provided that the following constraints are satisfied:
\begin{align}
\sum_{d \in D_i} y_d = 1,\quad      &\forall i = 1 \ldots n  \label{eq:eq3} \\[10pt] 
\sum_{d^{\prime} \in D_j : (d, d^{\prime}) \in \mathcal{E}_C} y_{d^{\prime}}  \geq y_d \quad &\forall d \in \mathcal{V}_C, \nonumber \\[-15pt] 
                                                                                       &\forall j=1 \ldots n : A_j  \in \mathcal{N}_{\alpha(d)}, j \neq \alpha(d) \label{eq:eq4}
\end{align}
Constraints~(\ref{eq:eq3}) ensure that exactly one path per train is selected (each agent can only assign one value to the variable it controls). Constraints~(\ref{eq:eq4}) state that, if path $d$ is selected and it is associated to agent $\alpha(d)$, then each neighbouring agent in $\mathcal{N}_{\alpha(d)}$ must select a path that is compatible with $d$.

By exploiting CPLEX as solver for this ILP formulation of the agent coordination in the dec-rtRTMP, we are able to find all possible solutions to a given problem instance, and among them, to identify the optimal ones as well.
%and among these to also identify the optimal solution.

\section{DSA}
\label{sec:suppl_dsa}

\begin{figure*}[!t]
    \centering
    \includegraphics[width=\textwidth]{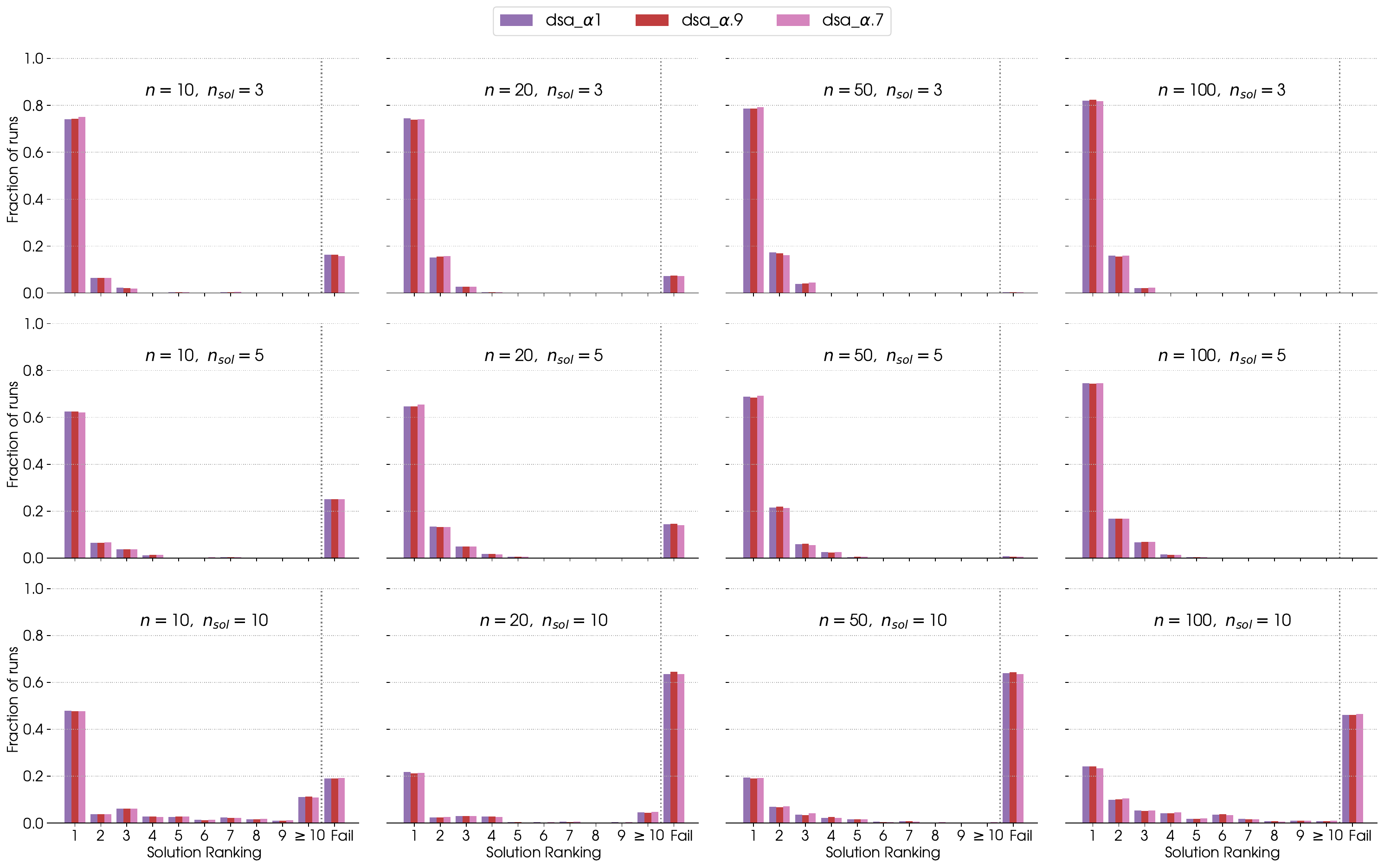}
    \caption{Ranking of the solutions given by DSA algorithms. Data are grouped by the type of agent, by number of agents $n$ and by minimum number of solutions $n_{sol}$ we require the problem instance to have. Bars with label ``$1$'' represent the fraction of executions that converged to an optimal solution. Bars with label ``$\geq 10$'' represent the fraction of executions that converged to a solution in position greater than 10 in the ranking. Bars with label ``Fail'' represent the fraction of executions that exceeded the upper bound of $10^5$ iterations.}
    \label{fig:suppl_dsa_ranking}
\end{figure*}

\begin{figure*}[!t]
    \centering
    \includegraphics[width=\textwidth]{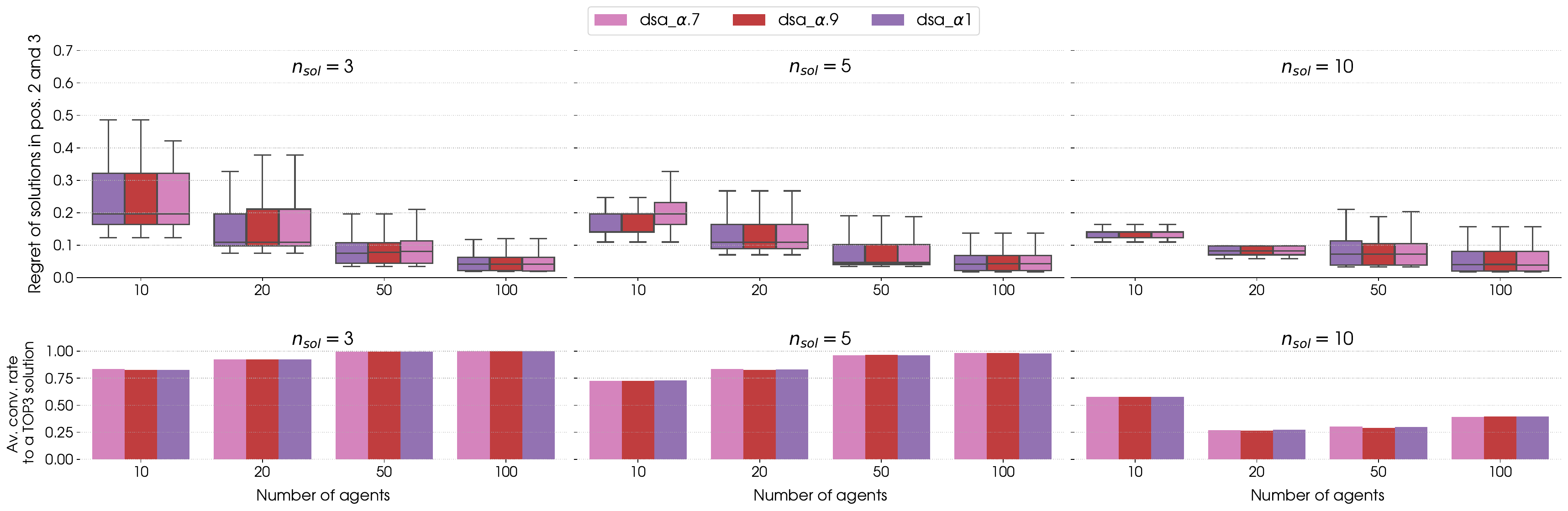}
    \caption{Top: Regret (percentage loss from the optimal solution value) of the solutions given by DSA algorithm in position 2 or 3 of the ranking. Data are grouped by type of agent, by number of agents $n$ and by minimum number of solutions $n_{sol}$ we require the problem instance to have.
    Bottom: Average convergence rate to a top-3 solution.}
    \label{fig:suppl_dsa_regret}
\end{figure*}

\begin{figure*}[!t]
    \centering
    \includegraphics[width=\textwidth]{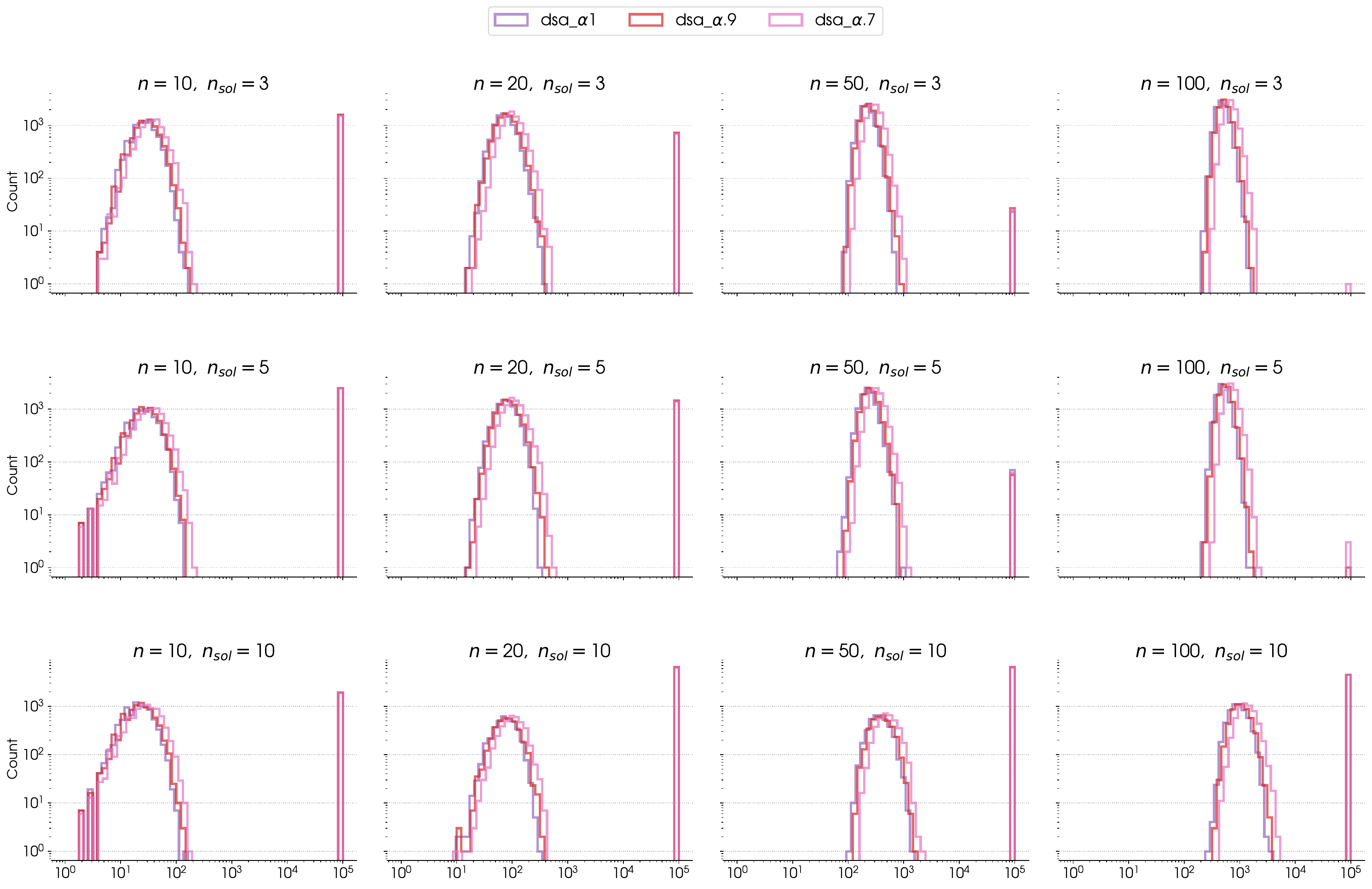}
    \caption{Distribution of convergence times (in terms of number of iterations) per type of agent on problem instances grouped by number of agents $n$ and by minimum number of solutions $n_{sol}$ we require the problem instance to have. For each of the 100 problem instances characterised by $n$ and $n_{sol}$, we performed $100$ executions of DSA algorithm for each agent type. Each execution has an upper bound of $10^5$ iterations, beyond which it fails.}
    \label{fig:suppl_dsa_convTime}
\end{figure*}

We run the DSA algorithm with $\alpha=1, 0.9, 0.7$ and evaluate it performance as in the main text. See Figures \ref{fig:suppl_dsa_ranking}, \ref{fig:suppl_dsa_regret} and \ref{fig:suppl_dsa_convTime}.

%%%%%%%%%%%%%%%%%%%%%%%%%%%%%%%%%%%%%%%%%%%%%%%%%%%%%%%%%%%%%%%%%%%%%%%%

\end{document}